\documentclass[pra,aps,twocolumn,superscriptaddress,showpacs]{revtex4}
\usepackage{hyperref}
\usepackage{bm}
\usepackage{epsfig}
\usepackage{amssymb}
\usepackage{graphicx}

\usepackage{amsmath}
\usepackage{epsfig}

\begin{document}
\title{Gaussian multipartite bound information}
\author{Ladislav Mi\v{s}ta, Jr.}
\affiliation{Department of Optics, Palack\' y University, 17.
listopadu 12,  771~46 Olomouc, Czech Republic}
\affiliation{School of Physics and Astronomy, University of St.
Andrews, North Haugh, St. Andrews, Fife, KY16 9SS, Scotland}
\author{Natalia Korolkova}
\affiliation{School of Physics and Astronomy, University of St.
Andrews, North Haugh, St. Andrews, Fife, KY16 9SS, Scotland}

\date{\today}

\begin{abstract}
We demonstrate the existence of Gaussian multipartite bound
information which is a classical analog of Gaussian multipartite
bound entanglement. We construct a tripartite Gaussian
distribution from which no secret key can be distilled, but which
cannot be created by local operations and public communication.
Further, we show that the presence of bound information is
conditional on the presence of a part of the adversary's
information creatable only by private communication. Existence of
this part of the adversary's information is found to be a more generic
feature of classical analogs of quantum phenomena obtained by
mapping of non-classically correlated separable quantum states.
\end{abstract}


\maketitle

Quantum entanglement and secret classical correlations are
nonlocal information resources in the sense that they cannot be
created by local operations and classical communication (LOCC)
\cite{Werner_89} and local operations and public communication
(LOPC) \cite{Maurer_93,Renner_03}, respectively. The
distillability of these resources is the key prerequisite for
their utility in reliable communication. Indeed, imperfect
entangled states can sometimes be distilled \cite{Bennett_96b} by
LOCC into a maximally entangled singlet state - a perfect resource
for the faithful transmission of quantum states via teleportation
\cite{Bennett_93}. Remarkably, there are entangled states which
cannot be distilled to singlet form \cite{MHorodecki_98}. These
states, fittingly said to be ``bound entangled'', demonstrate a
fundamental property of entanglement, namely its irreversibility
under LOCC transformations \cite{RHorodecki_09}. More precisely,
bound entanglement requires singlets for creation but no singlets
can be distilled from it. Despite being nondistillable and hence
unusable directly for quantum communication, bound entanglement
can be activated \cite{Dur_00,Masanes_06} and can contain
distillable entanglement when tensored \cite{Shor_03} or mixed
\cite{Dur_01}.

Moving to the secret correlations, they appear in the scenario in
which two honest parties, Alice and Bob, and an adversary Eve,
share independent realizations of three random variables $A,B$ and
$E$, characterized by the probability distribution
$P\left(A,B,E\right)$. As with entanglement, Alice and Bob can
sometimes distill them by LOPC to a secret key, i.e., a common
string of random bits about which Eve has practically no
information. Inspired by the existence of bound entanglement,
Gisin and Wolf posed a question \cite{Gisin_00} as to whether
there are also nondistillable secret correlations, referred to as
bound information (BI), that act as a classical analogue to bound
entanglement. So far, a bipartite distribution has been found
containing BI only in the asymptotic limit \cite{Renner_03}, but
surprisingly, an example of tripartite BI was derived in
\cite{Acin_04}. A probability distribution $P(A,B,C,E)$ shared by
three honest parties Alice, Bob and Clare, and an adversary Eve,
carries tripartite BI if \cite{Acin_04}: (i) any pair of honest
parties cannot distill a secret key, even if they can collaborate
with the third party, and (ii) the distribution cannot be
distributed by LOPC, i.e., the reduced probability distribution
$P\left(A,B,C\right)$ cannot be distributed among the honest
parties if their public communication is constrained to contain at
most information of the variable $E$.

The example of BI from Ref.~\cite{Acin_04} is discrete. It is
obtained from a specific bound entangled state $\hat{\rho}_{ABC}$
of three two-level systems by measuring its purification
$|\psi\rangle_{ABCE}$,
$\hat{\rho}_{ABC}=\mbox{Tr}_{E}(|\psi\rangle\langle\psi|_{ABCE})$,
where $E$ labels Eve's purifying system, with measurements
$\hat{P}_{i}$, $i=A,B,C,E$ in the computational basis as
\cite{Acin_05}
\begin{equation}\label{P}
P(A,B,C,E)=\mbox{Tr}(\hat{P}_{A}\otimes\hat{P}_{B}\otimes\hat{P}_{C}\otimes\hat{P}_{E}|\psi\rangle\langle\psi|_{ABCE}).
\end{equation}
The distribution contains secret correlations across one splitting
of honest parties into two groups which guarantees that both (i)
and (ii) hold. The key step in the construction of BI
\cite{Acin_04}, as well as of other classical analogs
\cite{Bae_09,Prettico_11} of quantum phenomena, is the proof of
the absence of secret correlations across particular splittings.
The proof is based on the concept of intrinsic information which
is for the distribution $P(A,B,E)$ defined as \cite{Maurer_99}
\begin{eqnarray}\label{upperbound}
I_{AB\downarrow E}=\mathop{\mbox{min}}_{E\rightarrow
\tilde{E}}I_{AB|\tilde{E}},
\end{eqnarray}
where $I_{AB|\tilde{E}}$ is the conditional mutual information and
the minimization is performed over all channels
$E\rightarrow\tilde{E}$. Importantly, if $I_{AB\downarrow E}=0$
then Alice and Bob do not share secret correlations
\cite{Renner_03}. In the proofs of
Refs.~\cite{Acin_04,Bae_09,Prettico_11} a suitable channel
$E\rightarrow\tilde{E}$ is found erasing a part of Eve's
information which nullifies the conditional mutual information.
The intrinsic information then vanishes and there are no secret
correlations across the considered splitting. The new variable
$\tilde{E}$ represents a publicly communicated message used by the
honest parties for the LOPC preparation of the reduced
distribution of their variables. However, in order to create the
{\it entire} considered distribution including the original
variable $E$, it is desirable to know whether Eve's part of the
information, which is erased by the channel
$E\rightarrow\tilde{E}$, can also be established by LOPC or
whether some other resources are needed.

In this paper we demonstrate the existence of multipartite BI for
random Gaussian continuous variables. We construct a tripartite
Gaussian distribution containing secret correlations across only
one splitting which thus carries tripartite Gaussian bound
information (GBI). By using the Gaussian entanglement criteria
\cite{Werner_01,Giedke_01}, we prove the presence of GBI without
using the concept of intrinsic information. This approach also
reveals that the creation of the part of Eve's information, which
is erased by the channel $E\rightarrow\tilde{E}$, requires private
communication between honest parties across the splittings where
the distribution contains no secret correlations. Moreover, we
find that this information is interconnected with the BI as if it
is dropped the BI disappears. Finally, we give an example of a
discrete distribution where creation of Eve's information requires
private communication and we argue that a necessary prerequisite
for this property is the mapping of the type (\ref{P}) of a
non-classically correlated separable quantum state.

{\it Gaussian secret correlations}. We consider the
set of Gaussian states and measurements associated
with quantum systems with infinite-dimensional Hilbert state spaces.
$N$ such systems, which can be physically realized by $N$ light
modes, are described by $N$ pairs of position and momentum
quadrature operators, labeled by $\hat{x}_{j}$ and $\hat{p}_{j}$,
respectively, $j=1,2,\ldots,N$, satisfying the canonical
commutation rules $[\hat{x}_{j},\hat{p}_{k}]=i\delta_{jk}$.
$N$-mode quantum states can be represented by the Wigner function
of $2N$ real variables and Gaussian states possess a Gaussian
Wigner function. Any $N$-mode Gaussian state $\hat{\rho}$ is
therefore fully characterized by a vector
$\bar\xi=\mbox{Tr}(\hat{\rho}\hat{\xi})$ of phase-space
displacements, where
$\hat{\xi}=\left(\hat{x}_{1},\ldots,\hat{x}_{N},\hat{p}_{1},\ldots,\hat{p}_{N}\right)^{T}$,
and by a covariance matrix (CM) $\gamma$ with elements
$\gamma_{jk}=2\mbox{Re}\mbox{Tr}[\hat{\rho}(\hat{\xi}_{j}-\bar{\xi}_{j})(\hat{\xi}_{k}-\bar{\xi}_{k})]$.

A Gaussian measurement consists of a set of Gaussian operators
obtained by all possible displacements of a fixed Gaussian state
and normalized such that the completeness relation is satisfied.
Incorporating Gaussian measurements and Gaussian states into the
mapping (\ref{P}), we can construct Gaussian distributions with
unique cryptographic properties. In the case of a bipartite state
$\hat{\rho}_{AB}$ the mapping can, e.g., yield a distribution
$P(A,B,E)$ with distillable secret correlations. Necessary and
sufficient conditions for secure-key distillability are not known,
yet nevertheless one can show \cite{Csiszar_78,Assche_04}, that a
Gaussian distribution $P\left(A,B,E\right)$ is distillable if
\begin{eqnarray}\label{lowerbound}
\mbox{max}\left(\Delta I_{DR},\Delta I_{RR}\right)>0,
\end{eqnarray}
where $\Delta I_{DR}=I_{AB}-I_{AE}$ and $\Delta
I_{RR}=I_{AB}-I_{BE}$ are differences of mutual information
\cite{Shannon_48}, $I_{AB}$ between Alice and Bob and $I_{AE}$
$(I_{BE})$ between Alice (Bob) and Eve.

The mapping of entanglement onto distillable secret correlations
is best exemplified by the two-mode squeezed vacuum state. It is a
Gaussian entangled state $|\tau(m)\rangle_{AB}$ of two modes $A$
and $B$ described by the CM
$\tau_{AB}(m)=\omega_{AB}(m)\oplus\left[\omega_{AB}(m)\right]^{-1}$
with
\begin{eqnarray}\label{omegaTMSV}
\left[\omega_{AB}(m)\right]_{ij}=\sqrt{m^2+\delta_{ij}-1},\quad
i,j=1,2,
\end{eqnarray}
where $m=\cosh(2r)$ ($r>0$ is the squeezing parameter). By measuring the position quadratures
on modes $A$ and $B$ (with outcomes $x_{A}$ and $x_{B}$) the state is mapped onto the Gaussian
distribution
\begin{eqnarray}\label{PTMSV}
P_{AB}(x_{A},x_{B})&=&(\pi)^{-1}e^{-\left(x_{A},x_{B}\right)\left[\omega_{AB}(m)\right]^{-1}\left({x_{A}
\atop x_{B}}\right)},
\end{eqnarray}
for which $I_{AB}=\log_{2}\left[\cosh\left(2r\right)\right]$. As
the state itself is pure, Eve is completely uncorrelated with
Alice and Bob. Consequently, $I_{AE}=I_{BE}=0$ and the global
distribution contains distillable secret correlations according to
the criterion (\ref{lowerbound}). A closer look at the
distribution further shows that it plays the role of a
continuous-variable analog of a basic unit of discrete secret
correlations, a secret bit \cite{Acin_05}, defined as a
probability distribution satisfying $P(A,B,E)=P(A,B)P(E)$ and
$P(A=B)=1$. Indeed, not only does the distribution satisfy the
first condition, but it also approaches the continuous-variable
analog $P_{AB}(x_{A},x_{B})\propto\delta(x_A-x_{B})$ of the second
condition with increasing $r$.

{\it Construction of GBI}. We construct GBI by mapping of a bound
entangled Gaussian state of three modes $A,B$ and $C$ with the CM
given in Eq.~(17) of Ref.~\cite{Mista_09}. To construct GBI from
the state we need to find its purification. The separability
criterion \cite{Werner_01} reveals that the state can be
decomposed into the product
$|\tau[\cosh(2r)]\rangle_{AC}|0\rangle_{B}$ of the two-mode
squeezed vacuum state in modes $A$ and $C$, and the vacuum state
$|0\rangle_{B}$ in mode $B$, and displacements
\begin{eqnarray}
\hat{x}_{\alpha}\rightarrow\hat{x}_{\alpha}'&=&\hat{x}_{\alpha}+\alpha_{x}q_x,\label{xdisplacements}\\
\hat{p}_{\alpha}\rightarrow\hat{p}_{\alpha}'&=&\hat{p}_{\alpha}+\alpha_{p}q_p,\quad
\alpha=A,B,C,\label{pdisplacements}
\end{eqnarray}
where $-A_{x}=C_{x}=A_{p}=C_{p}=-1/2$, $B_{x}=B_{p}=1$. The
classical displacements $q_x$ and $q_p$ obey a Gaussian
distribution ${\cal
P}(q_x,q_p)=\mbox{exp}[-(q_{x}^2+q_{p}^2)/(4x)]/(4\pi x)$, where
$x=(e^{2r}-1)/2$. Hence, we construct a suitable purification as
\begin{eqnarray}\label{psi}
|\psi\rangle_{ABCE}&=&\int\sqrt{{\cal
P}(q_x,q_p)}\hat{D}_{A}\left(\delta_{A}\right)\hat{D}_{C}\left(\delta_{C}\right)|\tau(m)\rangle_{AC}
\nonumber\\
&&\hat{D}_{B}\left(\delta_{B}\right)|0\rangle_{B}\left|q_x\right\rangle_{E_1}^{(x)}\left|q_p\right\rangle_{E_2}^{(p)}dq_xdq_p,
\end{eqnarray}
where $|\,.\,\rangle^{(x)}$ $\left(|\,.\,\rangle^{(p)}\right)$ is
a position (momentum) eigenstate, $m=\cosh(2r)$ and
$\hat{D}_{\alpha}\left(\delta_{\alpha}\right)=\mbox{exp}(\delta_{\alpha}\hat{a}_{\alpha}^{\dag}-\delta_{\alpha}^{\ast}\hat{a}_{\alpha})$
is the displacement operator with
$\delta_{\alpha}=(\alpha_{x}q_{x}+i\alpha_{p}q_{p})/\sqrt{2}$. The
purification (\ref{psi}) has the CM $\Gamma=X\oplus
\left(X\right)^{-1}$, where
\begin{eqnarray}\label{X}
X=\left(\begin{array}{ccccc}
a & 2x & b & 2x & \frac{1}{2}\\
2x & c & -2x & 4x & -e^{2r}\\
b & -2x & a & -2x & e^{2r}-\frac{1}{2}\\
2x & 4x & -2x & 4x & -2x\\
\frac{1}{2} & -e^{2r} & e^{2r}-\frac{1}{2} & -2x & y\\
\end{array}\right)
\end{eqnarray}
with $a=\cosh(2r)+x$, $b=\sinh(2r)-x$, $c=1+4x$ and
$y=e^{2r}\left(2e^{2r}-1\right)/\left[2\left(e^{2r}-1\right)\right]$.
By measuring the position quadratures on all modes of the
purification we get a Gaussian distribution
\begin{eqnarray}\label{Pi}
\Pi\left(\eta\right)&=&\left(\pi\right)^{-\frac{5}{2}}e^{-\eta^{T}\left(X\right)^{-1}\eta},
\end{eqnarray}
where $\eta=(x_A,x_B,x_C,x_{E_{1}},x_{E_{2}})^{T}$ is the vector
of measurement outcomes and $X$ is the classical covariance matrix
(CCM) (\ref{X}). The distribution contains secret correlations
across just one splitting and therefore contains multipartite GBI.

First we prove the absence of secret correlations across the
$B-(AC)$ splitting. Assume Alice and Clare privately draw two
random variables $z_{A}$ and $z_{C}$ from a Gaussian distribution
with CCM $\omega_{AC}[\cosh(2r)]$, Eq.~(\ref{omegaTMSV}). They
also draw a third variable $x_{E_{1}}$ from a Gaussian
distribution with variance $\langle x_{E_{1}}^{2}\rangle=2x$ and
send it through a public channel to Bob. He privately generates a
random variable $z_{B}$ obeying a Gaussian distribution with
variance $\langle z_{B}^{2}\rangle=1/2$. Alice, Bob and Clare
displace their variables as in Eq.~(\ref{xdisplacements}), where
we have performed the replacements
$\hat{x}_{\alpha}(\hat{x}_{\alpha}')\rightarrow
z_{\alpha}(x_{\alpha})$, $q_x\rightarrow x_{E_{1}}$.
 The variables $x_A,x_B$ and $x_C$ obey the reduced
distribution $\Pi(x_A,x_B,x_C)$. As the distribution was created
by the public communication of $x_{E_1}$ across the $B-(AC)$
splitting, the distribution (\ref{Pi}) indeed contains no secret
correlations across the splitting.

Similarly we can show that there are no secret correlations across
the $C-(AB)$ splitting. For this we again use the separability
criterion \cite{Werner_01} and decompose the underlying state into
the product $|\tau(m)\rangle_{AB}|0\rangle_{C}$, where
\begin{eqnarray}\label{m}
m=\frac{1+2(e^{4r}-e^{2r})}{2e^{2r}-1},
\end{eqnarray}
and the displacements (\ref{xdisplacements}) and
(\ref{pdisplacements}) with $A_{x}=-A_{p}=1/(2y)$,
$B_{x}=B_{p}=-e^{2r}/y$, $C_{x}=C_{p}=(1-e^{-2r})$, and the
displacements $q_x$ and $q_p$ obeying the Gaussian distribution
$\tilde{\cal
P}(q_x,q_p)=\mbox{exp}[-\left(q_x^2+q_p^2\right)/y]/(\pi y)$. The
decomposition tells us how to establish the reduced distribution
$\Pi(x_A,x_B,x_C)$ by LOPC with respect to the $C-(AB)$ splitting.
Initially, Alice and Bob draw privately two random variables
$z_{A}$ and $z_{B}$ from a Gaussian distribution with CCM
$\omega_{AB}(m)$, Eq.~(\ref{omegaTMSV}), where $m$ is given in
Eq.~(\ref{m}), and Clare privately generates a random variable
$z_{C}$ obeying a Gaussian distribution with variance $\langle
z_{C}^{2}\rangle=1/2$. Alice and Bob also draw a third variable
$x_{E_{2}}$ from a Gaussian distribution with variance $\langle
x_{E_{2}}^{2}\rangle=y/2$ and send it to Clare through a public
channel. All the participants then perform displacements
(\ref{xdisplacements}), where we have performed the replacements
$\hat{x}_{\alpha}(\hat{x}_{\alpha}')\rightarrow
z_{\alpha}(x_{\alpha})$, $q_x\rightarrow x_{E_{2}}$, and the
coefficients $\alpha_{x}$ are given below Eq.~(\ref{m}). The
variables $x_A,x_B$ and $x_C$ follow the reduced distribution
$\Pi(x_A,x_B,x_C)$, which we distributed by the public
communication of $x_{E_2}$ across the $C-(AB)$ splitting.
Consequently, the distribution (\ref{Pi}) contains no secret
correlations across the splitting.

\begin{figure}
\centerline{\psfig{width=7.5cm,angle=0,file=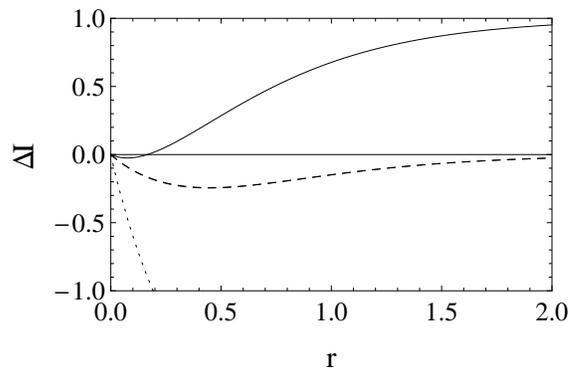}}
\caption{Information differences $\Delta I_{RR}$ (solid curve) in Eq.~(\ref{DeltaI}); $\Delta I_{RR}'$ (dotted
curve) and $\Delta I_{DR}'$ (dashed curve) for CCM (\ref{X}) with
respect to $A-(BC)$ splitting.}\label{figure}
\end{figure}

{\it Activating GBI}. As the distribution (\ref{Pi}) contains no
secret correlations across the $B-(AC)$ and $C-(AB)$ splittings,
no two honest parties can establish a secret key even with the
help of the third one. The distribution, however, cannot be
created by LOPC as it contains secret correlations across the
$A-(BC)$ splitting. They are not detected by the criterion
(\ref{lowerbound}) (see dotted and dashed curves in
Fig.~\ref{figure}) but can be ``activated'' by allowing Bob and
Clare to perform the joint operation
$x_{B,C}\rightarrow\left(x_{B}\pm x_{C}\right)/\sqrt{2}$. The
presence of secret correlations can be seen from the reduced
distribution of Alice's, Bob's and Eve's variables. The
corresponding information difference $\Delta I_{RR}$ arising in
the condition (\ref{lowerbound}) then can be expressed as
\begin{eqnarray}\label{DeltaI}
\log_{2}\sqrt{\frac{4-e^{-2r}-11e^{2r}+20e^{4r}-20e^{6r}+16e^{8r}}{2-8e^{2r}+10e^{6r}+4e^{8r}}}
\end{eqnarray}
and it is plotted against the squeezing parameter $r$ in Fig.~\ref{figure}.
The figure and numerics reveal, that $\Delta I_{RR}>0$ for $r>r_{\Delta I_{RR}=0}\doteq0.166$.
Therefore Alice and Bob can distill a secret key using the reverse
reconciliation protocol \cite{Grosshans_03}. This would be
impossible without the distribution (\ref{Pi}) to have secret
correlations across the $A-(BC)$ splitting which accomplishes the
construction of GBI.

{\it Specific adversary features}. The absence of secret
correlations could be alternatively proved using the intrinsic
information (\ref{upperbound}). If Eve uses the channel
$(x_{E_1},x_{E_2})\rightarrow x_{E_{1}} (x_{E_{2}})$, we get
$I_{B(AC)|E_{1}}=0$ ($I_{C(AB)|E_{2}}=0$). Then
$I_{B(AC)\downarrow E_1E_2}=0$ ($I_{C(AB)\downarrow E_1E_2}=0)$
follows and the distribution (\ref{Pi}) contains no secret
correlations across $B-(AC)$ ($C-(AB)$) splitting. It is, however,
advantageous to use the formalism of our previous proof as it
unveils the nonlocal nature of the part of Eve's information which
is erased by the channel, and its interconnection with the BI
carried by the distribution (\ref{Pi}). More precisely, it shows
first, that for $r>\tilde{r}\doteq0.284$ both of Eve's variables
are indispensable for the presence of the BI as dropping either of
them results in its disappearance. Indeed, the dropping of
$x_{E_1}$ causes the distribution $\Pi(x_A,x_B,x_{E_2})$ to give
$\Delta I_{DR}>0$ for $r>0.156$ and the secret key can be
distilled between Alice and Bob. Likewise, if $x_{E_2}$ is dropped
and $r>\tilde{r}$, Alice and Clare can distill a secret key with
the help of Bob (See Appendix~\ref{Appendix A} for the proof). In
the remaining interval $\tilde{r}\geq r>r_{\Delta I_{RR}=0}$ the
condition (\ref{lowerbound}) is not fulfilled if $x_{E_2}$ is
discarded and we cannot decide about the presence of BI.
Nevertheless, Alice and Bob can create a new variable
$x_{A}-x_{B}/2$ which, together with the variables $x_{C}$ and
$x_{E_1}$, obeys a distribution satisfying $\Delta I_{DR}>0$ for
$r>0$. Consequently, after discarding $x_{E_2}$ from the
distribution (\ref{Pi}) secret correlations across both the
$A-(BC)$ and $C-(AB)$ splittings are present and the property (i)
of BI is no longer guaranteed.

Let us now focus on the remaining variable $x_{E_2}$ ($x_{E_{1}}$)
which is erased in the proof of the absence of secret
correlations across the $B-(AC)$ ($C-(AB)$) splitting. Can honest
parties also publicly announce this variable by using only LOPC across the
splitting? If this was the case then it would be possible to create the
distribution with GBI (\ref{Pi}) just from secret correlations
with respect to the $A-C$ or $A-B$  splitting in the form of the
distribution (\ref{PTMSV}). Surprisingly, we answer the question
in the negative. Namely, the variable can be expressed as
\begin{eqnarray}\label{xEj}
x_{E_j}=-A_{p}z_{A}-B_{p}z_{B}-C_{p}z_{C}+ex_{E_{k}}+\chi_{E_{j}},
\end{eqnarray}
where $j=2$, $k=1$ ($j=1$, $k=2$) and
$e=(-1/2)(A_{x}A_{p}+B_{x}B_{p}+C_{x}C_{p})$. The parameters
$A_{x,p},B_{x,p},C_{x,p}$ are given below
Eq.~(\ref{pdisplacements}) (Eq.~(\ref{m})), and the random
variable $\chi_{E_2}$ ($\chi_{E_1}$) is uncorrelated with the
other variables and obeys a Gaussian distribution with variance
$\langle\chi_{E_{2}}^{2}\rangle=1/(8x)$
($\langle\chi_{E_{1}}^{2}\rangle=1/(2y)$). The variable
(\ref{xEj}) contains the term $A_{p}z_{A}+B_{p}z_{B}+C_{p}z_{C}$
involving variables belonging to both parts of the $B-(AC)$
($C-(AB)$) splitting. Assume, that the variable $x_{E_2}$
($x_{E_1}$) is known publicly and the parties share the reduced
distribution $\Pi(x_{A},x_{B},x_{C},x_{E_{1}})$
($\Pi(x_{A},x_{B},x_{C},x_{E_{2}})$), which can be completely
created by LOPC across the splitting. The honest parties can then
turn the ``nonlocal'' term into (distillable) secret correlations
by combining locally with respect to the $B-(AC)$ ($C-(AB)$)
splitting the variable $x_{E_2}$ ($x_{E_1}$) with their variables
(See Appendix \ref{Appendix B} for the proof). As it is impossible
to create secret correlations by LOPC the variable $x_{E_2}$
($x_{E_1}$) cannot be announced publicly by LOPC and private
communication across the $B-(AC)$ ($C-(AB)$) splitting is needed
for this task.

We have shown for the distribution (\ref{Pi}), that although it
does not contain secret correlations across a certain splitting,
the creation of the whole of Eve's information requires private
communication across the splitting. This property obviously cannot
be obtained by mapping of the type (\ref{P}) of pure states or
classically correlated separable states, which are diagonal in the
local product basis \cite{Piani_08} and product in the Gaussian
scenario \cite{Giorda_10}, but one can get it by mapping of some
non-classically correlated separable states which is the case of
the state used for construction of GBI. Interestingly, one can
find the property also for some other probability distributions
constructed via the mapping of such a state. For example, for the
discrete probability distribution (8) from Ref.~\cite{Bae_09}
derived from a fully separable state \cite{Cubitt_03} exhibiting
non-classical correlations \cite{Chuan_12}, Alice and Bob do not
share secret correlations. However, to create the distribution
Alice and Bob have to reveal to Eve, i.e., publicly announce, the
results of privately tossed, independent fair coins, but only if
the results differ. In those cases for which Alice and Bob do not
share their results with Eve, i.e., their results are the same, a
shared secret bit is established between them. This is clearly
impossible with LOPC and a private channel is required to decide
when to share their results with Eve.

{\it Conclusions}. We derived a distribution carrying tripartite
GBI. The entire distribution cannot be constructed by LOPC solely
from secret correlations shared by a certain pair of honest
parties because the creation of a part of Eve's information
requires private communication with the third honest party.
Moreover, this part of Eve's information is found to be
inextricably interconnected with the presence of the BI. As a
similar nonlocal part of Eve's information can also be traced for
some other classical analogs of quantum phenomena, an open
question arises about to which extent the interconnection between
this information and the presence of the phenomenon is general.
Our results reveal the nontrivial nature of the classical
information resources needed for formation of some classical
analogs of quantum phenomena and show that they are linked to the
quantum resources not only in the form of entanglement but also in
the form of separable non-classical correlations.

L. M. thanks to J. Fiur\'a\v{s}ek and R. Tatham for discussions.
The research has been supported by the EU FET-Open grant COMPAS,
No. 212008 and GACR Project No. P205/12/0694. N. K. is grateful to
the Alexander von Humboldt Foundation.

\appendix

\section{Indispensability of the variable $x_{E_2}$ for the presence of bound
information}\label{Appendix A}

In this appendix we show that if the variable $x_{E_2}$ is dropped
from the distribution (10) and $r>\tilde{r}\doteq0.284$, Alice and
Clare can distill a secret key with the help of Bob. Indeed, if
Bob publicly communicates his variable $x_{B}$ to Alice and Clare,
they can create new variables $x_{A}+g x_{B}$ and $x_{C}+h x_{B}$.
The distribution of these variables and the variables $x_{B}$ and
$x_{E_1}$ held by Eve, where $g$ and $h$ are chosen such that
$\Delta I_{DR}$ is maximized, gives $\Delta I_{DR}>0$ for
$r>\tilde{r}\doteq0.284$. Consequently, Alice and Clare can
distill a secret key with the help of Bob.

\section{Secret correlations appear if the erased
adversary's information is known publicly}\label{Appendix B}

In this appendix we prove, that if the parties share the reduced
distribution $\Pi(x_{A},x_{B},x_{C},x_{E_{1}})$
($\Pi(x_{A},x_{B},x_{C},x_{E_{2}})$) and the variable $x_{E_2}$
($x_{E_1}$) is publicly known, then the honest parties share
secret correlations across the $B-(AC)$ ($C-(AB)$) splitting.

Assume first, that all the participants hold the publicly known
variable $x_{E_{2}}$ and they also hold the reduced distribution
$\Pi(x_{A},x_{B},x_{C},x_{E_{1}})$. Then Alice and Clare can
create a new variable $x_{A}'=(x_{A}+x_{C})/2-x_{E_{2}}$. The
joint distribution of the variables $x_{A}',x_{B},x_{E_{1}}$ and
$x_{E_2}$ then satisfies $\Delta{I}_{DR}>0$ if $r>r'\doteq0.38$
and therefore contains (distillable) secret correlations with
respect to the $A-B$ splitting according to the criterion (3). If,
on the other hand, the honest parties create new variables
$x_{A}''=(x_{A}+x_{C})/2$ and $x_{B}''=x_{B}+x_{E_2}$, the
distribution of the variables $x_{A}'',x_{B}'',x_{E_{1}}$ and
$x_{E_2}$ yields $\Delta{I}_{DR}>0$ for $r<r''\doteq0.549$ and
consequently there are (distillable) secret correlations with
respect to the same splitting also for $r\leq r'$. As the latter
secret correlations have been established by local operations with
respect to the $B-(AC)$ splitting which cannot create secret
correlations, we have to conclude, that if the parties share the
distribution $\Pi(x_{A},x_{B},x_{C},x_{E_{1}})$ and know the
variable $x_{E_2}$, then the honest parties share secret
correlations across the $B-(AC)$ splitting.

Assume now, that the parties share the reduced distribution
$\Pi(x_{A},x_{B},x_{C},x_{E_{2}})$ and the publicly known variable
$x_{E_1}$. They can then prepare new variables
$\bar{x}_{A}=e^{2r}x_A+x_{B}/2$ and
$\bar{x}_{C}=x_{C}+ye^{-2r}x_{E_1}$ which together with the
variables $x_{E_1}$ and $x_{E_2}$ obey a distribution satisfying
$\Delta{I}_{DR}>0$ for $r>0$ across the $A-C$ splitting. Because
the latter secret correlations have been created by local
operations with respect to $C-(AB)$ splitting which cannot create
secret correlations, there are secret correlations across the
$C-(AB)$ splitting if the parties share the distribution
$\Pi(x_{A},x_{B},x_{C},x_{E_{2}})$ and know the variable
$x_{E_1}$.

\end{document}